\documentclass[conference]{IEEEtran}
\IEEEoverridecommandlockouts
\usepackage{cite}
\usepackage{verbatim}
\usepackage{amsmath,amssymb,amsfonts,bm}
\usepackage[linesnumbered,ruled,vlined]{algorithm2e}
\usepackage{bbm}
\usepackage{enumitem}
\usepackage{amsthm}
\usepackage{algpseudocode}
\usepackage{textcomp}

\usepackage{float}
\usepackage[font=small]{caption}
\usepackage{xcolor}
\usepackage{wrapfig}
\usepackage{hyperref}
\usepackage[font=footnotesize,labelformat=simple]{subcaption}
\usepackage[protrusion=true,expansion=true]{microtype}
\pdfoutput=1

\usepackage{caption}
\usepackage{multirow}
\usepackage{graphicx}
\everymath{\displaystyle}

\newtheorem{example}{Example}
\newtheorem{remark}{Remark}

\newcommand{\calC}{\mathcal{C}}

\newcommand{\calD}{\mathcal{D}}

\newcommand{\calM}{\mathcal{M}}

\newcommand{\calK}{\mathcal{K}}

\newcommand{\bfU}{\mathbf{U}}

\newcommand{\bfD}{\mathbf{D}}

\newcommand{\bfV}{\mathbf{V}}

\newcommand{\bfW}{\mathbf{W}}

\newcommand{\bfT}{\mathbf{T}}

\newcommand{\bfc}{\mathbf{c}}

\newcommand{\bfP}{\mathbf{P}}
\newcommand{\bfQ}{\mathbf{Q}}

\usepackage[compact]{titlesec}
\hypersetup{draft}

\def\BibTeX{{\rm B\kern-.05em{\sc i\kern-.025em b}\kern-.08em
    T\kern-.1667em\lower.7ex\hbox{E}\kern-.125emX}}
    
    \makeatletter 
\newcommand\semiHuge{\@setfontsize\semiHuge{22.7}{28.38}}
\makeatother

\begin{document}


\title{A Reinforcement Learning-Based Approach to Graph Discovery in D2D-Enabled Federated Learning}

\author{
\IEEEauthorblockN{Satyavrat Wagle, Anindya Bijoy Das, David J. Love, and Christopher G. Brinton}

\IEEEauthorblockA{Elmore Family School of Electrical and Computer Engineering, Purdue University}
}

\maketitle

\begin{abstract}
Augmenting federated learning (FL) with direct device-to-device (D2D) communications can help improve convergence speed and reduce model bias through rapid local information exchange. However, data privacy concerns, device trust issues, and unreliable wireless channels each pose challenges to determining an effective yet resource efficient D2D structure. In this paper, we develop a decentralized reinforcement learning (RL) methodology for D2D graph discovery that promotes communication of non-sensitive yet impactful data-points over trusted yet reliable links. Each device functions as an RL agent, training a policy to predict the impact of incoming links. Local (device-level) and global rewards are coupled through message passing within and between device clusters. Numerical experiments confirm the advantages offered by our method in terms of convergence speed and straggler resilience across several datasets and FL schemes.
\end{abstract}

\section{Introduction}\label{intro}

Federated Learning (FL) has become a popular approach for global machine learning (ML) model construction across a set of distributed edge devices. The standard operation of FL consists of a coordinating server periodically aggregating models trained locally at the edge devices on their respective local datasets. One of the fundamental challenges in FL is the presence of non-i.i.d data distributions across participating devices, which can slow convergence speed and result in global model bias~\cite{fed_to_fog}. These issues are exacerbated when some devices can only communicate their model updates to the server intermittently, e.g., due to poor channel conditions.

A recent trend of work has considered mitigating these issues through augmenting FL with device-to-device (D2D) communications in relevant network settings, e.g., wireless sensor networks~\cite{drl_multiuser_wsn}. In D2D-enabled FL, short-range information exchange is employed to reduce the tendency of devices to overfit on their locally collected datasets~\cite{wang2021device}. However, there are two factors which have a strong impact on the efficacy of such procedures: (i) \textit{inter-device trust and privacy concerns}, which may prevent data sharing between specific device pairs, possibly for certain data classes; (ii) \textit{D2D wireless condition variations}, which impacts communication efficiency and can result in intermittent data transmission failures.

In this paper, we ask: \textit{How can we facilitate discovery of an effective D2D structure for FL systems taking these factors into account?} To answer this, we propose a reinforcement learning (RL)~\cite{sutton2018} methodology for identifying links between devices that maximize a reward measuring FL performance and communication efficiency. In its decentralized form, device-specific policies (i.e., agents) can learn to independently predict these links through low-overhead message passing without complete exposure of local data distributions.



\textbf{Related work.} A few studies have explored bias reduction in FL models through D2D information exchange. For example, they have considered offloading of (i) partial data sets to compensate for heterogeneous computation capabilities across devices \cite{fed_to_fog}, (ii) data to devices which are estimated to contribute more to system performance \cite{wang2021device}, and (iii) unlabelled data for decentralized pseudo-labelling \cite{iotmal}. Our work, by contrast, considers D2D graph discovery to jointly optimize expected learning improvement and communication reliability.

Other works have employed RL to address similar issues. For example, \cite{flrl_dev_sel,flrl_dev_sel2} train policies at the server to select devices for aggregation that reduce the bias of the system model. In our work, by contrast, we leverage RL for D2D communication procedures. In addition, all of the above studies assume a centralized decision making system, which exposes additional device information to the network. To the best of our knowledge, a methodology to allow for \textit{device level decision-making} in the presence of inter-device trust constraints and variable communication channels has not been studied.


\subsection*{Summary of Contributions}
\begin{itemize}
    \item We propose a decentralized RL methodology for cooperative discovery of an efficient wireless communication graph for a D2D-enabled FL system (Sec. \ref{sec:graph_formation}). In our scheme, devices act as RL agents, training local policies for link formation and engaging in data/reward sharing.
    \item Our RL reward modeling and message passing procedure results in a D2D communication structure that (i) encourages reliable communication of impactful information (ii) in the presence of variable network conditions, while (iii) maintaining privacy requirements.
    \item We evaluate our method by conducting experiments on established datasets and FL schemes (Sec. \ref{sec:results}). Our method shows substantial improvement over baselines in terms of convergence speed, reliability of D2D communication, and robustness against the presence of stragglers.
\end{itemize}


\label{sec:network_model}
\begin{figure}[t!]
    \centering
    \includegraphics[width=0.8\columnwidth]{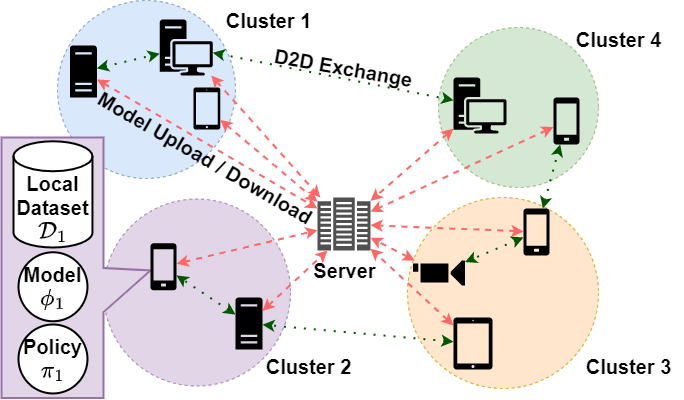}
    \caption{System model for D2D-enabled FL. In our RL-based approach for graph discovery,where each device acts as a learning agent.}
    \label{fig:sys_arch}\vspace{-5mm}
\end{figure}

\section{System Model and Problem Formulation}\label{sec:system_model}\vspace{-1mm}

In this section, we first detail our models for the wireless network and learning processes. Then, we formulate the D2D exchange graph discovery problem.

\subsection{Network and Learning Models}\label{sec:learning_model}
We consider an FL system over a network of client devices $\calC = \{ \calC_1, \calC_2, \dots, \calC_N\}$, hence $|\calC| = N$; which are regularly aggregated at server $\mathcal{S}$. Each device $\calC_i$ has access to a local dataset $\mathcal{D}_i$ and a local model $\phi_i^t \in \mathbb{R}^p$ where $p$ is the number of parameters, which is updated over the training period $t\in[0, T]$ to minimize a local cost function which is detailed in Section \ref{sec:learning_task}. Local models $\{\phi_i^t\}_{i \in \mathcal{C}}$ are aggregated every $\tau_a$ time steps at server $\mathcal{S}$ to obtain a global model $\phi_G^t$, which is broadcasted to all devices $\calC_i \in \mathcal{C}$ for further local training.
Now we discuss the aspects of our model that hold significance in any FL task.
\subsubsection{Device-to-Device Communication}
We assume that D2D communication can be established among the devices in $\mathcal{C}$ in order to exchange a subset of their local data-points with each other prior to the learning task. Recent works such as \cite{skewscout} imply that D2D exchange of a small number of data-points that reduce the non-i.i.d skew in local datasets yields significant performance gains in a learning task. However, predicting such D2D links may not have a straightforward solution additional resource costs required to account for unreliable channels due to network topology.
To limit this cost, we allow every receiver $\calC_i$ to receive data-points from at most one other remote device $\calC_j$, resulting in at most one incoming edge per device over which it receives data. 
 
Now, the signal received at $\calC_i$ may suffer from attenuation due to channel conditions between $\calC_i$ and $\calC_j$, which is observed in the received signal strength (RSS) at $\calC_i$.
We consider a network architecture similar to \cite{semidecentralized}, which assumes D2D communication conducted using OFDMA. In our scenario, for simplicity, we assume similar noise power $\sigma^2$ across all channels and a constant rate of transmission $r$ between devices. Therefore, we can express the probability of unsuccessful transmission to $\calC_i$ from $\calC_j$  similar to \cite{semidecentralized} as \vspace{-2mm}
\begin{align}\label{eq:prob_drop}
     \bfP_D(i,j) = 1 - \exp \left(\frac{-(2^r - 1) \cdot \sigma^2}{\bfW_{i,j}} \right),
 \end{align}
where $\bfW \in \mathbb{R}^{N \times N}$, such that $\bfW_{i,j}$ defines the RSS at $\calC_i$ when it receives a signal from device $\calC_j$. 
Thus, $\bfP_D$ is a useful indicator of system performance, which we use to design the reward function, which is detailed in Sec. \ref{sec:reward_modelling}
\subsubsection{Clusters of Reliable Devices} 
A low probability of failure of D2D communication is crucial. Hence, from the perspective of a receiver, it is important to identify links to remote devices which enable this. We therefore partition the devices in $\mathcal{C}$ into $K$ disjoint clusters, where each device $\calC_i$ belongs to a cluster $k$, denoted by $\mathcal{K}_k$, such that devices within a cluster are capable of reliably communicating between themselves. We define a reliable cluster of devices $\mathcal{K}_k$ as one where for all pairs of devices $\calC_i,\calC_j \in \mathcal{K}_k$; $\bfP_D(i,j) \leq \alpha_D$ ; where $\alpha_D$ is a reliability threshold set by the user. Thus, the probability of failure between any two devices within a cluster will always be less than $\alpha_D$. We can now define two forms of D2D communication as \textbf{intra-cluster} and \textbf{inter-cluster} communication.


Now, in order to minimize data exchange over unreliable channels (i.e, inter-cluster communication), we define a budget $B(\mathcal{K}_k)$ for each cluster $\mathcal{K}_k$, which limits the total number of data-points \textbf{requested} over inter-cluster links. Thus, the number of data-points requested by devices in $\mathcal{K}_k$ from devices which are not in $\mathcal{K}_k$ can be at most $B(\mathcal{K}_k)$. For any $k = 1,2, \dots, K$. If $\bfQ_{j \rightarrow i}$ denotes the number of data-points requested by the receiver $\calC_i \in \calK_k$, we formally define this constraint as \vspace{-1.mm}
\begin{align}\label{eq:budget}
     \sum_{\calC_i \in \mathcal{K}_k} \sum_{\calC_j \notin \mathcal{K}_k}  \bfQ_{j \rightarrow i} \leq B(\mathcal{K}_k).
\end{align}

\vspace{-2mm}
\subsubsection{Trust between Devices} 
In D2D communication, another desired property is protection against privacy breaches. In other words, devices are prohibited from sharing sensitive data with other devices unless the receiver is trusted by the transmitter; for example, a camera equipped device may want to share images other than those of humans for privacy concerns, except with certain trusted devices. We encode this notion of trust in a device specific trust matrix which is denoted by $\bfT_i \in \mathbb{R}^{N \times L}$, where $L$ is the number of classes in the overall dataset $\mathcal{D} = \bigcup_{\calC_i \in \mathcal{C}} \mathcal{D}_i$. The entries of trust matrix $\bfT_i$ belong to the set $\{ 0, 1\}$ , given by\vspace{-1.5mm}
\begin{align} \label{eq:trust_matrix}
    \bfT_i[j,\ell] = \begin{cases}
        1 & \texttt{if $\calC_i$ trusts $\calC_j$ with class $\ell$}\\
        0 & \texttt{else}.
    \end{cases}
\end{align}
Thus, in our system model, we do not allow a device $\calC_i$ to transmit data-points of class $\ell$ to device $\calC_j$ if $\bfT_i[j,\ell] = 0$. Note that $\bfT_i[j,\ell] = 1$ does not imply that $\bfT_j[i,\ell] = 1$.



\subsubsection{Learning Task} \label{sec:learning_task}

Finally, once intelligent D2D data exchange has been conducted; the local model $\phi_i^t$ at each device is updated at every time step $t$ to achieve a local learning task, as described in Sec. \ref{sec:learning_model}. We consider a classification task, where each client $\calC_i$ has its own local data-set $\calD_i$ which consists of tuples $(d, \ell) \in \mathcal{D}_i$ where $d$ is the feature vector for any data-point and $\ell$ is the corresponding class. The performance of the local model $\phi_i^t$ depends on a loss function $\mathcal{L}(\phi_i^t, \mathcal{D}_i)$, where
\begin{align}
    \mathcal{L}(\phi_i^t, \mathcal{D}_i) = \sum_{(d,\ell) \in \mathcal{D}_i} \texttt{CELoss}(\phi_i^t, d,\ell)
\end{align}
where $\texttt{CELoss}$ is the Cross Entropy Loss between the predicted and ground truth classes. Now, in the FL setting, the goal of the system is to learn a global model $\phi_G^*$ such that
\begin{align}\label{eq:global_func}
    \phi_G^* = \underset{\phi \in \mathbb{R}^p}{\arg\min} \sum_{i = 1}^{|\mathcal{C}|} \mathcal{L}(\phi, \mathcal{D}_i) 
\end{align}
The optimal global model is expected to perform the classification task with high accuracy across the global data distribution $\mathcal{D} = \bigcup_{i \in \mathcal{C}} \mathcal{D}_i$. 

\subsection{Graph Discovery Problem Formulation}
\label{sec:problem_form}

As local $\phi_i^t$ models are updated, they are expected to diverge over the training iterations between aggregation \cite{fedavg}, resulting in slow convergence of $\phi_G^*$. Studies such as \cite{robust} have shown that this effect is more pronounced when the data-sets across devices are non i.i.d. Our aim is to enable faster convergence of $\phi_G^*$ through cooperative D2D information exchange by improving local data diversity. 

Here we define class-distribution vector of local data at client $\calC_i$ as $\bfD_i \in \mathbb{R}^{L}$, where $L$ is the total number of classes in global dataset $\mathcal{D}$ and $\ell$-th entry of $\bfD_i$ is the number of local data-points of class $\ell$ available in client $\calC_i$.
Now, due to the nature of stochastic gradient descent, which is used to optimize local models $\phi_i^t$, the number of data-points required to create a noticeable improvement in the local data diversity (hence, in the learning task), must be above a certain threshold. 
We denote this threshold by $\bfc_i \in \mathbb{R}^L$, where any entry $\bfc_i[\ell]$ indicates the threshold for the corresponding class $\ell$ for client $\calC_i$. 
$\bfc_i[\ell]$ can be user defined, as different scenarios may limit the number of data-points that can be shared over wireless channels. Now, we take into account the skew of classes across devices by first defining a diversity threshold $\hat{L}$, which is set by the user. We ensure that each device $\calC_i$ has at least $\hat{L}$ classes available in their local dataset after D2D exchange by imposing the following constraint:\vspace{-2.mm}
\begin{align}\label{eq:data_diversity}
    (\sum_{\ell = 1}^L |\bfD_i[\ell] \geq \bfc_i[\ell]|) \geq \hat{L} .
    \vspace{-2.mm}
\end{align} 
Therefore, in short, our goal in this paper is to identify the links over the set of devices $\mathcal{C}$ that improve the diversity metric $\sum_{\ell = 1}^L |\bfD_i[\ell] \geq \bfc_i[\ell]|$ for every $\calC_i$, while ensuring that the requirement specified in (\ref{eq:data_diversity}) is fulfilled to create an optimal communication graph. Note that this needs to be optimized subject to the constraints mentioned in Sec. \ref{sec:network_model} which include (i) allowing at most one incoming edge for every device $\calC_i$, (ii) maintaining a minimum received signal strength (RSS) for every transmission, (iii) not allowing the transmission of a prohibitively large number data-points between two different clusters as shown in \eqref{eq:budget} and (iv) abiding by the notions of trust defined between the devices in \eqref{eq:trust_matrix}.

\section{Proposed Methodology}\label{sec:graph_formation}

\begin{figure}[t]
    \centering
    \includegraphics[width=0.95\columnwidth]{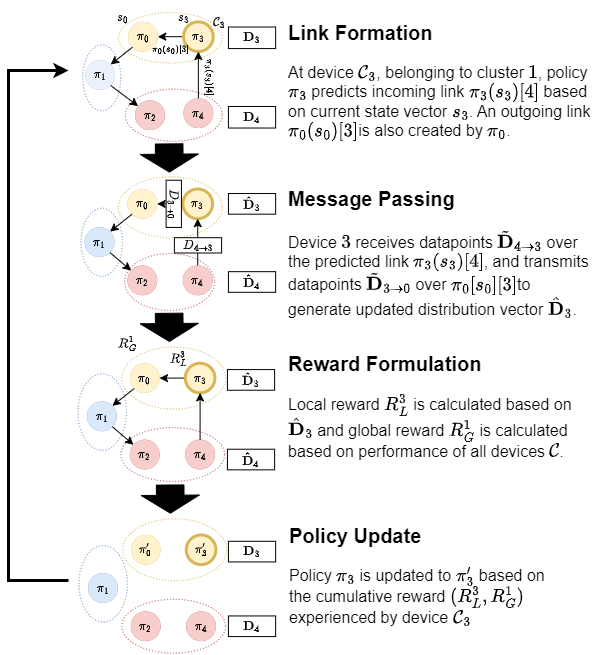}
    \caption{The intelligent graph discovery process iteratively improves local policies in a decentralized manner by updating them such that information exchange over the predicted links maximize a system-wide performance metric. }
    \label{fig:model}
    \vspace{-7mm}
\end{figure}

In this section, we discuss our proposed method, which discovers an optimal D2D communication graph over the devices $\calC$ to solve the problem established in Sec. \ref{sec:problem_form}. In order to do so, we use a decentralized RL framework, which trains a set of local policies $\pi = \{\pi_i \in \mathbb{R}^{N}\}_ {i \in [1,N]}$ that are used to jointly predict links among devices. These links aim for reliable sharing of salient data-points which reduce the skew of datasets at each client, while obeying system constraints. 


We use the Q-Learning framework \cite{sutton2018} for training the policy $\pi$, where each policy training episode consists of four steps. The first one is \textbf{link formation}, where the set of policies $\pi$ predict a set of links over the graph of devices $\calC$. The next one is the \textbf{message passing} procedure, which decides the payload of data-points to be transmitted over the predicted edges.  After that, the \textbf{reward formulation} step calculates the utility of links predicted by $\pi$ based on a reward function. The final step is the \textbf{policy update} which updates $\pi$ in an iterative manner, based on the experienced rewards and leads to the discovery of an optimal graph. The steps are discussed below in details.
%

%


\subsection{Link Formation}\label{sec:link_formation}
\vspace{-1.mm}

In this step, we first define the state at device $\calC_i$ as $s_i^t = \{\bfW_{i,j} : \calC_j \in \mathcal{C}\} \in \mathbb{R}^{N}$ indexed by $t$, where $\bfW_{i,j}$ is the RSS at device $\calC_i$ while receiving a signal from device $\calC_j$ as defined in (\ref{eq:prob_drop}). The set of possible states is given by $\calM$, and $|\calM| = S$. We also define an experience buffer at each device $\calC_i$ as $\psi_i \in \mathcal{R}^{S \times N \times 2}$, which is used to store computed rewards and will be discussed in Sec. \ref{sec:policy_update}. $\psi_i[s,j,0]$ is the total reward and $\psi_i[s,j,1]$ counts the frequency, respectively, for all times that $\pi_i$ selects a link from device $\calC_j$ when in state $s$ over the policy training process. Each device $\calC_i$ predicts an incoming edge from $\calC_j$ using its local policy $\pi_i$ and $s_i^t$ with probability 
\begin{align}\label{eq:prob_edge}
    \pi_i(s_i^t)[j] = \frac{\exp(\frac{\psi_i[t,j,0]}{\psi_i[t,j,1]})}{\sum_{k \in \mathcal{C}} \exp(\frac{\psi_i[t,k,0]}{\psi_i[t,k,1]})}
\end{align}
Once links are predicted for all receiving nodes, information is exchanged over the edges of the directed graph over $\calC$ based on the message passing algorithm, which we describe next.

\subsection{Message Passing}\label{sec:message_passing}
\vspace*{-5mm}
\begin{algorithm}
\caption{D2D Message Passing} \label{algo:message_passing}
\begin{algorithmic}[1]
    \State \textbf{Given :} Receiver node $\calC_i$, Transmitter node $\calC_j$, current state $s$, policy $\pi$
    \State Transmitter $\calC_j$ calculates the available data for exchange as $\bfV_{j \rightarrow i}$ using (\ref{eq:avl_at_tx}) and shares it with receiver $\calC_i$
    \State Receiver $\calC_i$ computes the data diversity vector $\bfD_i$ according to (\ref{eq:data_diversity})
    \State Receiver $\calC_i$ calculates the required data vector $\bfQ_{j \rightarrow i}$ using (\ref{eq:req_at_dst})
    \State Transmitter $\calC_j$ selects local data-points to add to transmission buffer $\bfU_{j \rightarrow i}$ using (\ref{eq:tx_buffer}) and transmits them to receiver $\calC_i$.
\end{algorithmic}
\end{algorithm}
\vspace*{-2mm}
The goal of the message passing algorithm is to select data-points to be transmitted over a link, while maintaining notions of trust between the transmitter and the receiver. We also ensure that the D2D exchange does not result in the transmitter having a more biased local data-set than before. The logic for the message passing algorithm is as follows.

 Let $\mathcal{N}_j$ be the set of devices requesting data-points from transmitter $\calC_j$ after the link formation step. $\calC_j$ first transmits the number of data-points that are available for sharing with all devices $\calC_i \in \mathcal{N}_j$ as a vector $\bfV_{j \rightarrow i} \in \mathbb{R}^L$. We calculate $\bfV_{j \rightarrow i}[\ell]$ as follows\vspace{-2.mm}
\begin{align}\label{eq:avl_at_tx}
\bfV_{j \rightarrow i}[\ell] = \mathbbm{1}_{\bfT_j[i,\ell] = 1}(\bfD_{j}[\ell] - \bfc_{j}[\ell])
\end{align}

The above equation ensures that transmitter $\calC_j$ only shares those data-points that are allowed by trust matrix $\bfT_j$. 

Upon receiving $\bfV_{j \rightarrow i}$, receiver $\calC_i$ forms a requirement vector $\bfQ_{j \rightarrow i} \in \mathbb{R}^{L}$, where $\bfQ_{j \rightarrow i}[\ell]$ is the number of data-points of class $\ell$ requested by $\calC_i$ from $\calC_j$ and is calculated as follows\vspace{-2.mm}
\begin{align}\label{eq:req_at_dst}
\bfQ_{j \rightarrow i}[\ell] = \begin{cases}
        \bfV_{j \rightarrow i}[\ell] , &\texttt{if } \bfc_{i}[\ell] - \bfD_i[\ell] \geq \bfV_{j \rightarrow i}[\ell] \\
        \bfc_{i}[\ell] - \bfD_i[\ell] , &\texttt{if }  0 < \bfc_{i}[\ell] - \bfD_i[\ell] < \bfV_{j \rightarrow i}[\ell] \\
        0 , &\texttt{if } \bfc_{i}[\ell] - \bfD_i[\ell] \leq 0
        \end{cases}\vspace{-2.mm}
\end{align}
The requirement vector $\bfQ_{j \rightarrow i}$ is then shared with transmitter $\calC_j$. Based on $\bfQ_{j \rightarrow i}$, $\calC_j$ selects data-points of class $\ell$ from $\mathcal{D}_j$ for all classes $\ell \in \mathcal{L}$ and forms a transmission buffer $\bfU_{j \rightarrow i} \in \mathbb{R}^L$. The transmission buffer $\bfU_{j \rightarrow i} \in \mathbb{R}^L$ contains the number of data-points that $\calC_j$ is \textbf{actually} able to share. This may differ significantly from $\bfV_{j \rightarrow i}$ due to the different demands $R_{j \rightarrow i'}$ made by all $i' \in \mathcal{N}_j$ devices. Note that, if the total demand is higher than what $\calC_j$ can afford to transmit, data-points are sent based on relative demand from each receiver $\calC_i\in \mathcal{N}_j$. We calculate transmission buffer $\bfU_{j \rightarrow i}$ as follows.\vspace{-1mm}
\begin{align}\label{eq:tx_buffer}
\bfU_{j \rightarrow i}[\ell] = \begin{cases}
    \bfQ_{j \rightarrow i}[\ell] , \;\; \texttt{if} \sum_{\calC_{i'} \in \mathcal{N}_j} R_{j \rightarrow i'}[\ell] \leq \bfD_{j}[\ell] - \bfc_j[\ell] \\
    \frac{\bfQ_{j \rightarrow i}[\ell]}{\sum\limits_{\calC_{i'} \in \mathcal{N}_j} \bfQ_{j \rightarrow i'}[\ell]} \cdot (\bfD_{j}[\ell] - \bfc_{j}[\ell]) , \;   \texttt{else}
\end{cases}\vspace*{-5mm}
\end{align}
Now, as transmission buffer $\bfU_{j \rightarrow i}$ drops packets with probability $P_D(i,j)$ as per (\ref{eq:prob_drop}), receiver $\calC_i$ is receives a buffer $\tilde{\bfD}_{j \rightarrow i}$, such that $\tilde{\bfD}_{j \rightarrow i}[\ell] \leq \bfU_{j \rightarrow i}[\ell]~\forall~\ell \in L$ and forms an updated class distribution vector $\hat{\bfD}_i$ as follows
\begin{align}\label{eq:final_diversity_vector}
\hat{\bfD}_i[\ell] = \bfD_i + \tilde{\bfD}_{j \rightarrow i} - \hspace*{-2mm}\sum_{k \in \mathcal{N}_i} \tilde{\bfD}_{i \rightarrow k} , j \sim \pi_i(s_i^t)
\end{align}
In our simulations, we model the expected number of received data-points $\tilde{\bfD}_{j \rightarrow i}$ as $\tilde{\bfD}_{j \rightarrow i}[\ell] = [1-P_D(i,j)]\bfU_{j \rightarrow i}[\ell]$.

\begin{figure}[t]
    \centering
    \includegraphics[width=0.9\columnwidth]{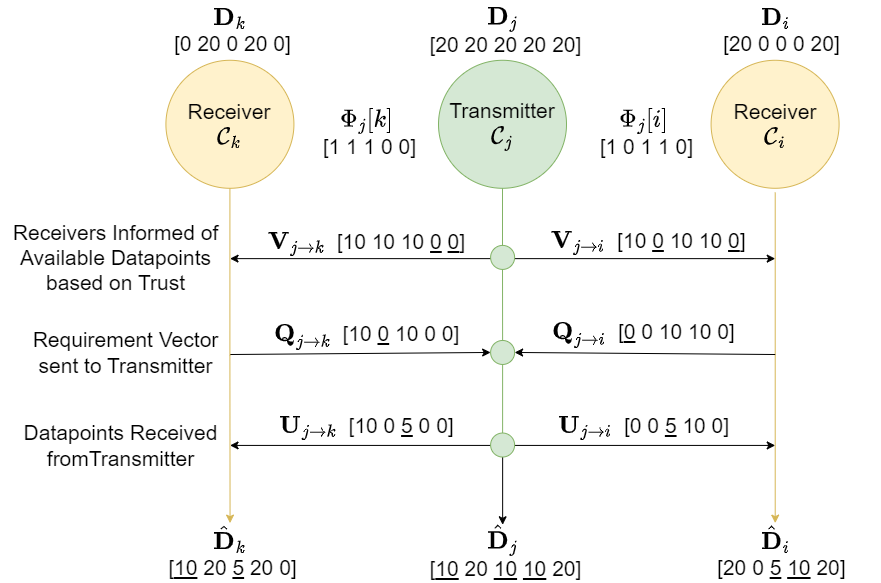}
    \caption{An example of message passing process for $\bfc_j[\ell] = 10 ~\forall~\ell$.}
    \label{fig:message_passing}\vspace{-6mm}
\end{figure}

The message passing algorithm is outlined in Algorithm \ref{algo:message_passing}.
\begin{example} 
{\normalfont 
In Fig. (\ref{fig:message_passing}), transmitter $\calC_j$ shares data-points only from trusted classes with receivers $\calC_i$ and $\calC_k$, while preserving enough for its own threshold constraints to be satisfied. Consider $\ell=3$, where the total demand $\bfQ_{j \rightarrow k} + \bfQ_{j \rightarrow i} = 20$, is greater than what is available at $\calC_i$ to share, which is $\bfV_{j \rightarrow i}[\ell] = \bfV_{j \rightarrow k}[\ell] = 10$. Here, the demand is split between both, so that $\tilde{\bfD}_{j \rightarrow k}[\ell] = \bfU_{j \rightarrow i}[\ell] = 5$ and $\hat{\bfD}_{j}[\ell] = 10$.
}
\end{example}

\begin{remark}
{\normalfont
Note that for any receiver $\calC_i$, the data distribution $\bfD_i$ is never fully exposed to a transmitter $\calC_j$, unless $\bfT_j[i,k] = 1 ~\forall ~ k$ (complete trust). Also, due to (\ref{eq:tx_buffer}), $\calC_j$ may want to share fewer data-points from a class $\ell$ with requesting devices, as $\calC_j$ must be left with at least $\bfc_j[\ell]$ data-points after each exchange.
}
\end{remark}
\begin{remark}
{\normalfont
Intuitively, the improved diversity of $\hat{\bfD_i}$ mitigates the detrimental effect of straggler devices \cite{wang2021device} within the system by ensuring that datapoints of any class are available at more devices. We explore this effect further in Sec. \ref{sec:strg_res}.
}
\end{remark}



\vspace{-1mm}
\subsection{Reward Modelling} \label{sec:reward_modelling}

Now, we use the updated class distribution vectors $\hat{\bfD_i}$ to formulate a reward structure for the system. This enables the policies to learn device specific requirements through a local reward, while also optimizing the system-wide metrics via a global reward. Therefore, the cumulative reward experienced by each device $\calC_i$ should take into consideration (i) performance of its local policy $\pi_i$, (ii) performance of other devices $\{\calC_j\}_{j \neq i, j\in \mathcal{C}}$, (iii) reliability of the received signal, as defined in (\ref{eq:prob_drop}) and (iv) the inter-cluster transmission, as defined in (\ref{eq:budget}).  

We now briefly discuss the parameters influencing cumulative reward. 
The local data diversity, as defined in (\ref{eq:data_diversity}), $\hat{\bfD_i} ~\forall~ i \in [1,N]$ should increase after D2D data exchange to improve convergence speed.  
Also, for a predicted link between $\calC_i$ and $\calC_j$, the probability of failed transmissions $P_D(i,j)$ as defined in (\ref{eq:prob_drop}), should be low in order to reliably receive signals over a selected edge. Also, as defined in (\ref{eq:budget}), the total number of data-points received via inter-cluster exchange must be less than the data budget $B(\mathcal{K}_k)$. Trust concerns are handled by the message passing algorithm in (\ref{eq:avl_at_tx}).In order to incorporate all of the above metrics, the overall reward must constitute a tradeoff, which is characterized by user defined weights $\alpha_1,\alpha_2,\alpha_3$. The reward consists of two components, a {\bf local reward} $r_L^{i}$ specific to device $\calC_i$, and a {\bf global reward} $r_G^{\calK_k}$ specific to cluster $\calK_k$. 

The local reward is independent of the system performance and captures the performance of the policy $\pi_i$ for device $\calC_i$ in terms of data diversity and reliability. In order to account for the data diversity requirement (\ref{eq:data_diversity}), for a given diversity threshold $\hat{L}$, we first define a score function $f:(\mathbb{R}^L,\mathbb{R}^L) \rightarrow \mathbb{R}$, which maps a diversity vector $\bfD_i$ and a set of threshold values $\bfc_i = \left[ \bfc_i[1] \;\; \bfc_i[2] \dots \bfc_i[L] \right]$ as follows \vspace*{-1mm}
\begin{align*}
    f(\bfD_i,\bfc_i) = \begin{cases}
         \sum_{\ell = 1}^L \mathbbm{1}_{\bfD_i[\ell] \geq \bfc_i[\ell]} &,\texttt{if} \; \left(\sum_{\ell = 1}^L \mathbbm{1}_{\bfD_i[\ell] \geq \bfc_i[\ell]} \right) \geq \hat{L}\\
         0 &, \texttt{otherwise}.
     \end{cases} \vspace*{-7mm}
\end{align*}

\vspace*{-1mm}
The utility function $f$ ensures that the predicted links satisfy the data diversity requirement in \eqref{eq:data_diversity}, by only resturning rewards if the condition is met. Thus, we define the local reward as \vspace*{-1mm}
\begin{align}
    r_L^i = \underbrace{\alpha_1 \cdot f(\hat{\bfD}_i,c)}_{\text{Data Diversity}} - \underbrace{\alpha_2 \cdot  (P_D(i,j))}_{\text{Reliability Maximization}} , j \sim \pi_i(s_i^t).
\end{align}
\vspace{-0.1 cm}

Next, the global reward $r_G^{\calK_k}$ captures the performance of the overall network, ensuring that all devices on average improve while cluster budget constraints are met. To that end, devices share their local $r_L^i$ rewards with other devices in the network. Budget constraints are found by obtaining the number of data-points received over inter-cluster links for device $\calC_{i'} \in \mathcal{K}_k$ as 
$    \tilde{Q}_{\mathcal{K}_k} = \sum_{i' \in \mathcal{K}_k}
        |\bfQ_{j \rightarrow i'}| $ where $ j \sim \pi_i(s_i^t)$ and $\calC_j \notin \mathcal{K}_k$.
We now define the global reward as
\vspace*{-2mm}
\begin{align}
    r_G^{\mathcal{K}_k} = \underbrace{\sum_{i \in C} \frac{r_L^i}{N}}_{\text{System Performance}} + \underbrace{\alpha_3 \cdot (B(\mathcal{K}_k) - \tilde{Q}_{\mathcal{K}_k})}_{\text{Cluster Budget}};
    \vspace*{-1mm}
\end{align}
The overall reward for a client $\calC_i \in \mathcal{K}_k$ is given by 
$    R_i^{\mathcal{K}_k} = r_L^i + \gamma \cdot r_G^{\mathcal{K}_k}$;
where the weighting term $\gamma$ governs the importance given to the overall performance of the system. If $\gamma$ is large, the devices tolerate a large reduction in local rewards if the global reward improves as a result, while a small $\gamma$ results in devices greedily optimizing their local rewards. Next, we discuss how the reward $R_i^{\mathcal{K}_k}$ is used to update local policy $\pi_i$.

\subsection{Policy Update}\label{sec:policy_update}

We use a decentralized multi-agent Q-Learning algorithm to update a policy $\pi_i$ in a state indexed by $s$, selected an edge from $\calC_j$ resulting in a reward $R_i^{\mathcal{K}_k}$ as follows
\begin{align*}
    \psi_i[s,j,0] = \psi_i[s,j,0] + R_i^c(t); \;\;\; \psi_i[s,j,1] = \psi_i[s,j,1]+1.
\end{align*}
This update increases the probability of a policy predicting links that maximize the experienced rewards, specified in (\ref{eq:prob_edge}).

\section{Simulation Results and Discussion}\label{sec:results}
In this section, we illustrate the advantages of our algorithm against baselines in terms of convergence speed, energy consumption, reliability of D2D communication, consistency in the presence of stragglers and delayed model aggregations.

\begin{figure*}[h!]	
	\centering
	\begin{subfigure}[t]{0.24\textwidth}
		\centering
		\includegraphics[width=0.99\linewidth,height=3.5cm]{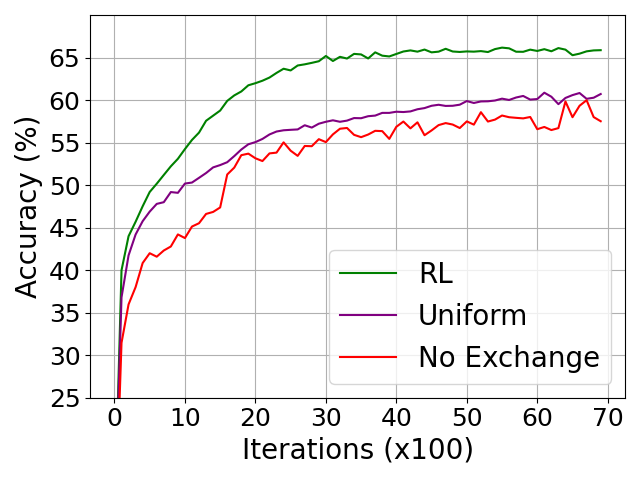}
            \vspace{-7mm}
		\caption{}\label{fig:2a}		
	\end{subfigure}
        \begin{subfigure}[t]{0.24\textwidth}
		\centering
            \includegraphics[width=0.99\linewidth,height=3.5cm]{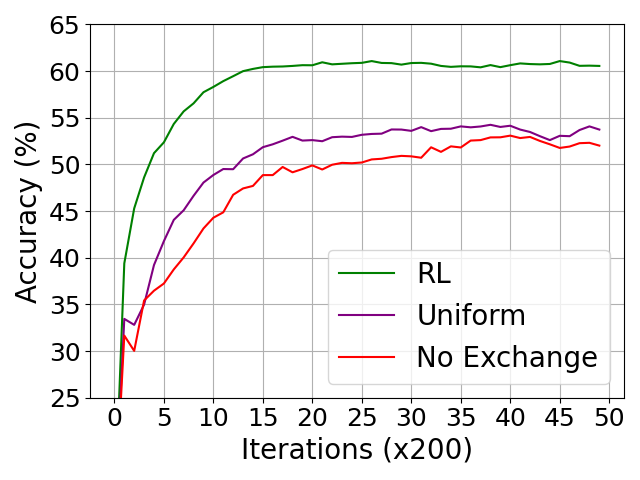}
            \vspace{-7mm}
		\caption{}\label{fig:2b}		
	\end{subfigure}
        \begin{subfigure}[t]{0.24\textwidth}
		\centering
            \includegraphics[width=0.99\linewidth,height=3.5cm]{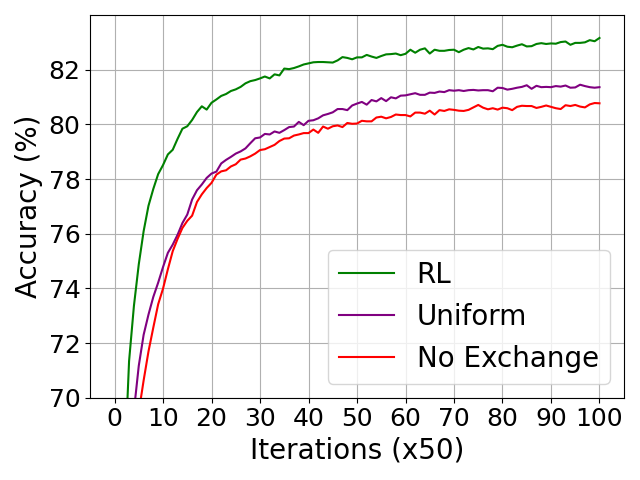}
            \vspace{-7mm}
		\caption{}\label{fig:2c}		
	\end{subfigure}
        \begin{subfigure}[t]{0.24\textwidth}
		\centering
		\includegraphics[width=0.99\linewidth,height=3.5cm]{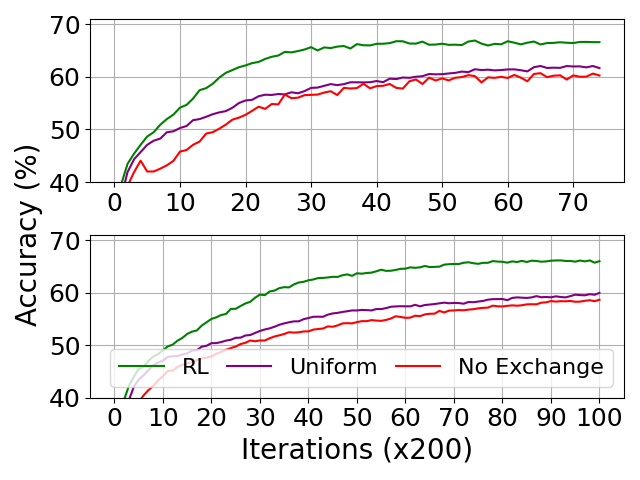}
            \vspace{-7mm}
		\caption{}\label{fig:2d}		
	\end{subfigure}
        \begin{subfigure}[t]{0.24\textwidth}
		\centering
		\includegraphics[width=0.99\linewidth,height=3.5cm]{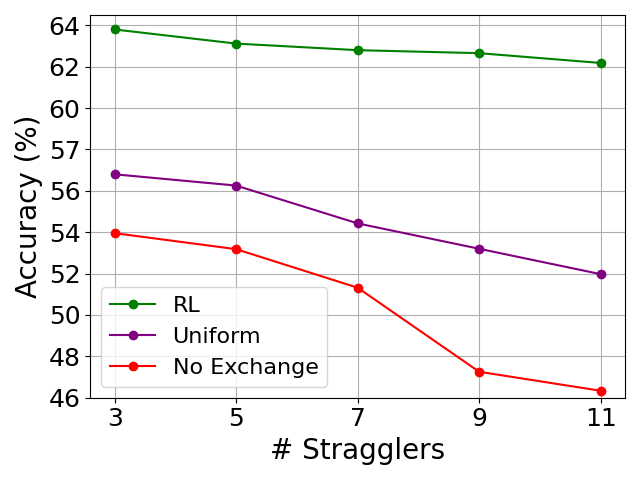}
            \vspace{-7mm}
		\caption{}\label{fig:2e}		
	\end{subfigure}
        \begin{subfigure}[t]{0.24\textwidth}
		\centering
            \includegraphics[width=0.99\linewidth,height=3.5cm]{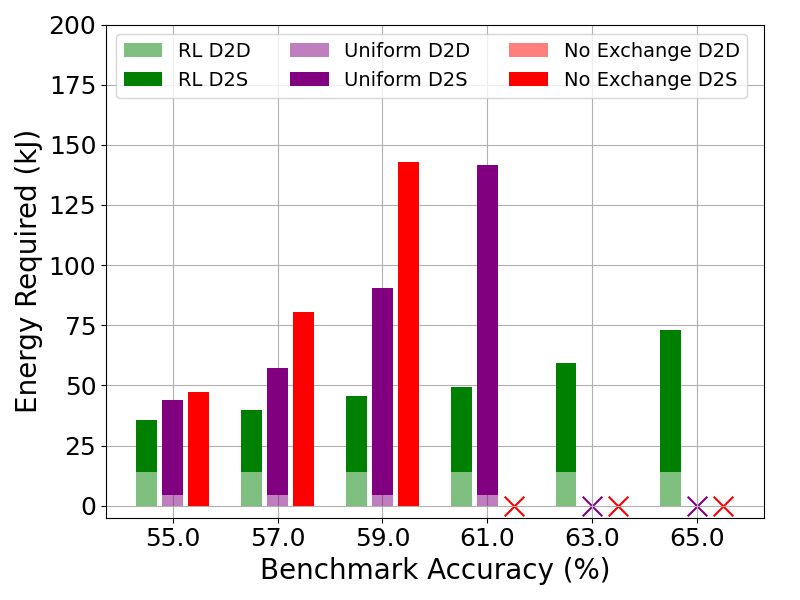}
            \vspace{-7mm}
		\caption{}\label{fig:2f}		
	\end{subfigure}
        \begin{subfigure}[t]{0.24\textwidth}
		\centering
		\includegraphics[width=0.99\linewidth,height=3.5cm]{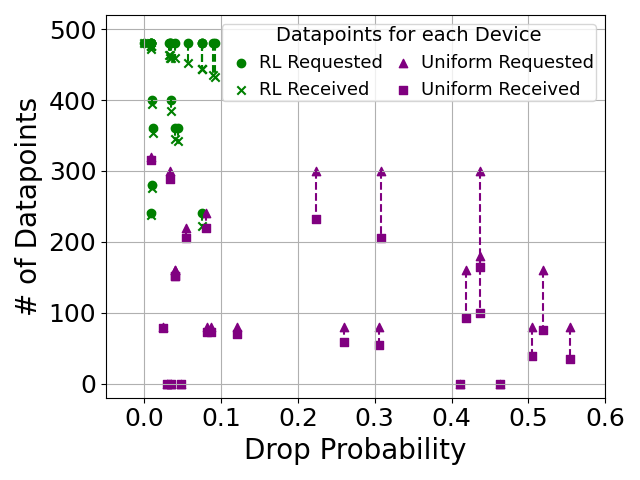}
            \vspace{-7mm}
		\caption{}\label{fig:2g}		
	\end{subfigure}
        \begin{subfigure}[t]{0.24\textwidth}
		\centering
		\includegraphics[width=0.99\linewidth,height=3.5cm]{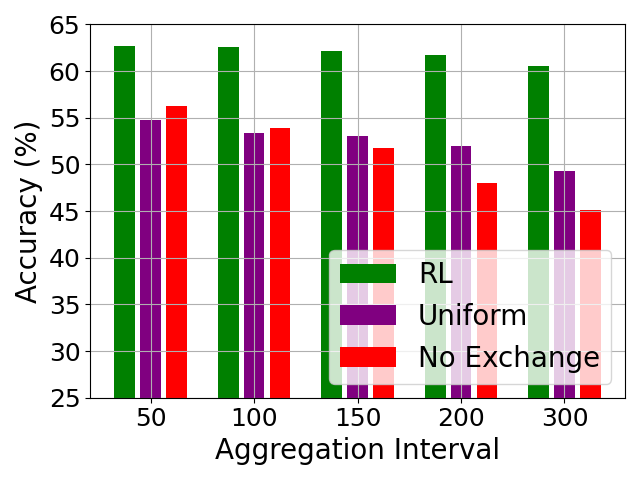}
            \vspace{-7mm}
		\caption{}\label{fig:2h}		
	\end{subfigure}
	\caption{Simulation results: Our method significantly improves performance over baselines for (a) RadioML ,(b) CIFAR10 and (c)FMNIST. It can be used to augment multiple existing federated learning algorithms such as (d) FedProx and FedSGD. It retains performance in the presence of stragglers (e) and consumes less energy to reach performance milestones (f). (g) Our method significantly improves the probability of successful D2D transmission. (h) The performance of our method remains relatively consistent over larger global aggregation intervals.
 }
 \label{fig:2}\vspace{-4.7mm}
\end{figure*}
\subsection{Experimental Setup}


We use the RadioML \cite{radioml}, CIFAR10 \cite{cifar10} and FashionMNIST \cite{fmnist} datasets for our evaluations. All datasets are split $80/20$ to obtain training and testing datasets respectively. We consider a network of $N = 25$ devices, and emulate non i.i.d training data across all devices. Each device has $990$ samples per device for RadioML, and $1200$ for the CIFAR10 and FashionMNIST from $4$ different classes. We use a convolutional neural network (CNN) as the FL model for RadioML and CIFAR10, and a fully connected network for FashionMNIST.


\subsection{Results and Discussion}

{\bf Performance on Various Datasets:} Now, we compare our algorithm on various datasets with the following baselines; (i) without data exchange and (ii) graphs generated using the Erd\H{o}s-Renyi model with uniform edge selection probability (denoted as ``uniform'') paired with the message passing algorithm (Alg. \ref{algo:message_passing}). We show this comparison  on RadioML, CIFAR10 and Fashion MNIST in Figs. \ref{fig:2a}, \ref{fig:2b} and \ref{fig:2c}, respectively. The results illustrate that D2D information exchange using our method improves the FL performance significantly over both of the competing scenarios. We emphasize that our approach finds a desirably structured D2D communication graph, resulting in considerable improvement of the FL performance over the ``uniform'' case irrespective of overall dataset. 

\vspace{1 mm}

{\bf Varying FL Schemes:} Next we apply our method to two other FL schemes: FedProx \cite{fedprox} and FedSGD \cite{fedavg}, and compare the performance in Fig. \ref{fig:2d}. We observe that our method significantly outperforms both baselines which indicates that it can be applied over different popular FL schemes without sacrificing performance gains. In FedProx, our method complements the proximal dissimilarity term by reducing model bias via D2D exchange. In FedSGD, gradient aggregation is also benefited by our method as the bias in the model gradient is reduced due to data similarity.

\vspace{1 mm}

{\bf Effect of Stragglers on Performance:} We now study the performance of our method in the presence of straggler devices \cite{wang2021device} in the FL system which do not participate in model aggregation. Thus, as the number of stragglers increases, fewer local models are aggregated. As each model is biased towards non-i.i.d local data, it reduces the accuracy of the global model.
\label{sec:strg_res}
In Fig. \ref{fig:2e}, we choose stragglers randomly from the devices and show that our method is more resilient to stragglers than the baselines. It indicates the ability of our method to share data that reduces the bias of the aggregated model as the data exchange allows the system to make up for the bias introduced by stragglers. This shows that our method is inherently robust to node failure and heterogenous communication capabilities. 



\vspace{1 mm}

{\bf Energy Consumption to reach Benchmarks:} Next, we conduct a simulation to compare the energy required by our method to achieve performance benchmarks with baselines. We use the wireless energy consumption model \cite{eed2d} to calculate the energy consumed for D2D and device-to-server (D2S) communication. In this simulation, we assume that the D2S distance is $3\times$ the average D2D distance. Fig. \ref{fig:2f} shows that our method uses significantly less energy to reach benchmarks as baselines despite the initial overhead due to D2D exchange. Note that in the ``uniform'' case, suboptimal links result in fewer data-points exchanged, resulting in lower D2D energy, but consequently significant higher D2S energy as a result.

\vspace{0.8 mm}
{\bf Reliability of D2D Performance:} In Fig. \ref{fig:2g}, we study D2D reliability in terms of the probability of successful transmission and the cluster budget.  We observe that our method consistently predicts links to reduce inter cluster communication while improving system performance. In practice, it results in a reduction of number of transmitted packets (which is not the case in ``uniform'' graphs), thus saves additional costs required to ensure successful transmission over unreliable channels.


\vspace{0.8 mm}
{\bf Change in Aggregation Interval:} Next, we observe the effect of various aggregation intervals $\tau_a$, or the frequency of local models synchronization. A low $\tau_a$ can result in faster convergence, but involves a larger overhead due to frequent D2S communication required for synchronization. Fig. \ref{fig:2h} shows that as $\tau_a$ becomes larger, our method outperforms the baselines by a considerable margin which indicates its resilience to delays in model aggregation, and a lower local model drift. Thus, a small initial overhead for our method results in significant gains in D2S overhead by retaining similar performance. 

\bibliographystyle{IEEEtran}
\bibliography{gbcm_v2.bib} 

\end{document}